\documentclass[12pt, onecolumn]{article}

\usepackage[dvipdfmx]{graphicx}
\usepackage[utf8]{inputenc}
\usepackage[T1]{fontenc}
\usepackage[numbers]{natbib}
\usepackage{amsmath}
\usepackage{amsthm}
\usepackage{amssymb}
\usepackage{algorithm}
\usepackage{algorithmic}
\usepackage{ecltree}
\usepackage{enumerate}
\usepackage{xcolor}
\usepackage{comment}
\usepackage{tikz}
\usepackage{wrapfig}
\usepackage{mathrsfs}
\usepackage{mathtools}
\usepackage[margin=25truemm]{geometry}
\mathtoolsset{showonlyrefs=true}

\usepackage{bm}
\usepackage{bbm}
\usepackage{float}
\usepackage{cases}
\usepackage{longtable}
\usepackage{diagbox}
\usepackage{booktabs}

\newtheorem{proposition}{Proposition}
\newtheorem{lemma}[proposition]{Lemma}
\newtheorem{theorem}[proposition]{Theorem}

\theoremstyle{definition}
\newtheorem{definition}[proposition]{Definition}

\newtheorem{remark}[proposition]{Remark}

\newcommand{\V}{\mathcal{V}}
\newcommand{\E}{\mathcal{E}}
\newcommand{\NAE}{\operatorname{NAE}}

\makeatletter

\@addtoreset{equation}{section}
\makeatother

\title{Closing the Complexity Gap for\\ Wheeler Graph Recognition on $d$-NFAs}
\author{Kengo Hashimoto}
\date{}

\allowdisplaybreaks
\begin{document}
\maketitle

\begin{abstract}
Wheeler graph recognition asks whether a directed graph whose arcs are labeled by symbols from a totally ordered alphabet admits a Wheeler order, a total order on its vertices compatible with the labels of arcs and the order of their tails.
A $d$-NFA is a nondeterministic finite automaton in which at most $d$ transitions with any given label leave each state.
For Wheeler graph recognition on $d$-NFAs, Alanko et al.\ gave a polynomial-time algorithm for $d \leq 2$, whereas Gibney and Thankachan proved \textsf{NP}-completeness for every fixed $d \geq 5$.
The computational complexity for $d = 3$ and $d = 4$ remained open.
We resolve both open cases by extending the reduction of Gibney and Thankachan with a new encoding of clauses of Not-All-Equal 3-SAT.
Specifically, we prove that recognition is \textsf{NP}-complete even for directed acyclic graphs over an ordered binary alphabet with a unique source reaching every vertex and total out-degree at most three at every vertex.
Consequently, Wheeler graph recognition on $d$-NFAs is \textsf{NP}-complete for every fixed $d \geq 3$.
\end{abstract}

\section{Introduction}
\emph{Wheeler graphs}, introduced by Gagie et al.~\cite{GMS20}, provide a common framework for extending efficient data structures based on the Burrows--Wheeler transform from strings to directed graphs whose arcs are labeled by symbols from a totally ordered alphabet.
A Wheeler graph admits a total order on its vertices, called a \emph{Wheeler order}, that is compatible with the labels of arcs and the order of their tails.
This order guarantees \emph{path coherence}: the set of vertices reachable from any interval of vertices by following paths spelling a fixed string is itself an interval.
This property enables exact pattern-matching queries to be processed by maintaining intervals rather than enumerating individual paths.
Therefore, \emph{Wheeler graph recognition}, which asks whether a given directed arc-labeled graph admits a Wheeler order, is a fundamental problem.

Previous work has investigated how the complexity of recognition depends on the maximum number of equally labeled arcs leaving a vertex.
For a positive integer $d$, a directed arc-labeled graph with a single source from which every vertex is reachable is called a \emph{$d$-NFA} if, for every vertex $v$ and symbol $a$, at most $d$ outgoing arcs from $v$ are labeled with $a$.\footnote{We regard a nondeterministic finite automaton (NFA) as its transition graph over a totally ordered finite alphabet $\Sigma$, omitting the accepting states because they do not affect Wheeler graph recognition.}
For $d \leq 2$, Alanko et al.~\cite{ADPP20} gave a polynomial-time algorithm that determines whether a given $d$-NFA is a Wheeler graph and constructs a Wheeler order when one exists.
Gibney and Thankachan~\cite{GT19} proved that recognition is \textsf{NP}-complete for every fixed $d \geq 5$, even for directed acyclic graphs (DAGs) over a binary alphabet.
The complexity of the two remaining cases, $d = 3$ and $d = 4$, was left open~\cite{ADPP20,Gib20}.

\subsection{Contribution}
In this paper, we resolve the remaining cases by proving the following stronger result,
which shows hardness under a bound of three on the total out-degree rather than the out-degree for each label.

\begin{theorem}
\label{thm:main}
Wheeler graph recognition is \textsf{NP}-complete even for DAGs whose arcs are labeled by symbols from an ordered binary alphabet, with maximum total out-degree at most three and a single vertex of in-degree $0$ from which every vertex is reachable.
\end{theorem}

Since every graph in this restricted class is a $3$-NFA, Wheeler graph recognition on $d$-NFAs is \textsf{NP}-complete for every fixed $d \geq 3$.

Our proof is based on a reduction from Not-All-Equal 3-SAT~\cite{Sch78}.
It builds on the \emph{betweenness gadget} and the use of successive copy layers in the reduction by Gibney and Thankachan~\cite{GT19}.
Our main contribution is a new encoding of clauses that enforces the not-all-equal condition while keeping the total out-degree of every vertex at most three.

The remainder of the paper is organized as follows.
Section~\ref{sec:preliminaries} briefly reviews Wheeler orders and NAE-3-SAT and presents the betweenness gadget introduced by Gibney and Thankachan~\cite{GT19}.
Section~\ref{sec:proof} proves Theorem~\ref{thm:main} by constructing a graph associated with an instance of NAE-3-SAT and establishing the correctness of the reduction.

\section{Preliminaries}
\label{sec:preliminaries}
For a non-negative integer $n$, let $[n] \coloneqq \{1, 2, \ldots, n\}$.

In the following, a \emph{graph} over a totally ordered alphabet $(\Sigma, \leq)$ is a tuple $G = (\V, \E, \ell)$, where $\V$ is the set of vertices, $\E \subseteq \V \times \V$ is the set of arcs, and $\ell \colon \E \to \Sigma$ is a mapping that assigns a label to each arc.

\begin{definition}[\cite{GMS20}]
\label{def:wheeler}
Let $G = (\V, \E, \ell)$ be a graph over $(\Sigma, \leq)$.
A \emph{Wheeler order on $G$} is a total order $\leq$ on $\V$ satisfying the following conditions (a) and (b).
\begin{enumerate}[(a)]
\item For any $u, v \in \V$, if $d^-(u) = 0$ and $d^-(v) > 0$, then $u < v$.
Namely, vertices with in-degree $0$ precede all vertices with non-zero in-degree.
\item For any $e_1 = (u_1, v_1) \in \E$ and $e_2 = (u_2, v_2) \in \E$, the following conditions (b1) and (b2) hold.
\begin{itemize}
\item[(b1)] If $\ell(e_1) < \ell(e_2)$, then $v_1 < v_2$.
\item[(b2)] If $\ell(e_1) = \ell(e_2)$ and $u_1 < u_2$, then $v_1 \le v_2$.
\end{itemize}
\end{enumerate}
The graph $G$ is called a \emph{Wheeler graph} if there exists a Wheeler order on $G$.
\end{definition}

Although the same symbol $\le$ is used for the total orders on both $\Sigma$ and $\V$, the context makes clear which is intended.

\begin{remark}
Note that condition~(b1) implies that all incoming arcs to any given vertex in a Wheeler graph must share the same label. Consequently, parallel arcs differing only in their labels can never exist in any Wheeler graph.
Also, parallel arcs with the same label can be replaced by a single arc without changing the set of valid Wheeler orders or increasing any out-degree.
We may therefore assume that the graph has no parallel arcs and represent it by an arc set $\E\subseteq\V\times\V$ and a label map $\ell\colon\E\to\Sigma$.
\end{remark}

Hereinafter, let $(\Sigma, \leq)$ be the totally ordered binary alphabet $\Sigma = \{0, 1\}$ with $0 < 1$.
We write a $0$-labeled arc and a $1$-labeled arc from $u$ to $v$ as $u \xrightarrow{0} v$ and $u \xrightarrow{1} v$, respectively.

\begin{definition}
Not-All-Equal 3-SAT (NAE-3-SAT) is a decision problem defined as follows.
An instance is a set of Boolean variables $x_1, \dots, x_n$ and a Boolean formula $\Phi = \bigwedge_{j = 1}^m \NAE(a_j, b_j, c_j)$,
where $a_j, b_j, c_j$ are literals and $\NAE(a_j, b_j, c_j) \coloneqq (a_j \lor b_j \lor c_j) \land (\bar{a}_j \lor \bar{b}_j \lor \bar{c}_j)$.
The problem asks whether there exists a truth assignment $\tau : \{x_1, \dots, x_n\} \to \{\mathsf{True}, \mathsf{False}\}$ satisfying $\Phi$, i.e., such that for every clause, the three literals are neither all $\mathsf{True}$ nor all $\mathsf{False}$.
NAE-3-SAT is well-known to be \textsf{NP}-complete~\cite{Sch78}.
\end{definition}

For three distinct vertices $u, v, w \in \V$, the \emph{betweenness constraint} $B(u, v, w)$ with respect to a total order $\le$ on $\V$ asserts that $v$ lies strictly between $u$ and $w$ in the order $\leq$; that is, either $u < v < w$ or $w < v < u$ holds.
Following Gibney and Thankachan~\cite{GT19}, we employ a dedicated gadget to enforce betweenness constraints as follows.

\begin{figure}[htpb]
\centering
\includegraphics[keepaspectratio,scale=0.8]{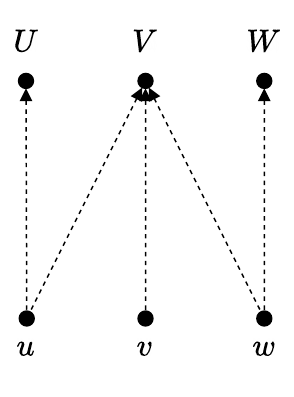}
\caption{A betweenness gadget for $B(u, v, w)$, where dashed lines represent $1$-labeled arcs.}
\label{fig:gadget}
\end{figure}

\begin{lemma}
\label{lem:gadget}
Let $u, v, w$ be three distinct vertices in a graph $G = (\V, \E, \ell)$.
The constraint $B(u, v, w)$ can be enforced in a Wheeler order by adding a \emph{betweenness gadget} consisting of three new vertices $U, V, W$ and the following five $1$-labeled arcs, as shown in Fig.~\ref{fig:gadget}:
\begin{align}
\mathcal{B}(u, v, w; U, V, W)
\coloneqq \{u \xrightarrow{1} U, u \xrightarrow{1} V, v \xrightarrow{1} V,
w \xrightarrow{1} V, w \xrightarrow{1} W\}, \label{eq:ihmnmt8ybwp6}
\end{align} 
where $U, V, W$ have no other incoming or outgoing arcs in the augmented graph.

More precisely, for a fixed total order on $\V$, the following equivalence holds in the graph augmented with the betweenness gadget:
the constraint $B(u, v, w)$ is satisfied if and only if there exists a relative order on $\{U, V, W\}$ such that Definition~\ref{def:wheeler}~(b1) and~(b2) hold for all pairs of arcs in $\mathcal{B}(u, v, w; U, V, W)$.
\end{lemma}

\begin{proof}
Without loss of generality, assume $u < w$.

First, suppose $B(u, v, w)$ is not satisfied; without loss of generality, let $v < u < w$.
For the arcs $v \xrightarrow{1} V$ and $u \xrightarrow{1} U$, the assumption $v < u$ and Definition~\ref{def:wheeler}~(b2) require $V \le U$.
On the other hand, for the arcs $u \xrightarrow{1} U$ and $w \xrightarrow{1} V$, the assumption $u < w$ and Definition~\ref{def:wheeler}~(b2) require $U \le V$.
These inequalities imply $U = V$, which contradicts $U \neq V$.
By symmetry, the case $u < w < v$ similarly leads to a contradiction.

Conversely, suppose $B(u, v, w)$ is satisfied, i.e., $u < v < w$.
It is directly verified that assigning the order $U < V < W$ satisfies Definition~\ref{def:wheeler}~(b1) and~(b2) for all pairs of arcs in $\mathcal{B}(u, v, w; U, V, W)$.
\end{proof}

\section{Proof of Theorem \ref{thm:main}}
\label{sec:proof}

The recognition problem belongs to \textsf{NP}
because one can verify in polynomial time whether a given total order on $\V$
is a Wheeler order.

\begin{figure}[htpb]
\centering
\includegraphics[keepaspectratio,scale=0.75]{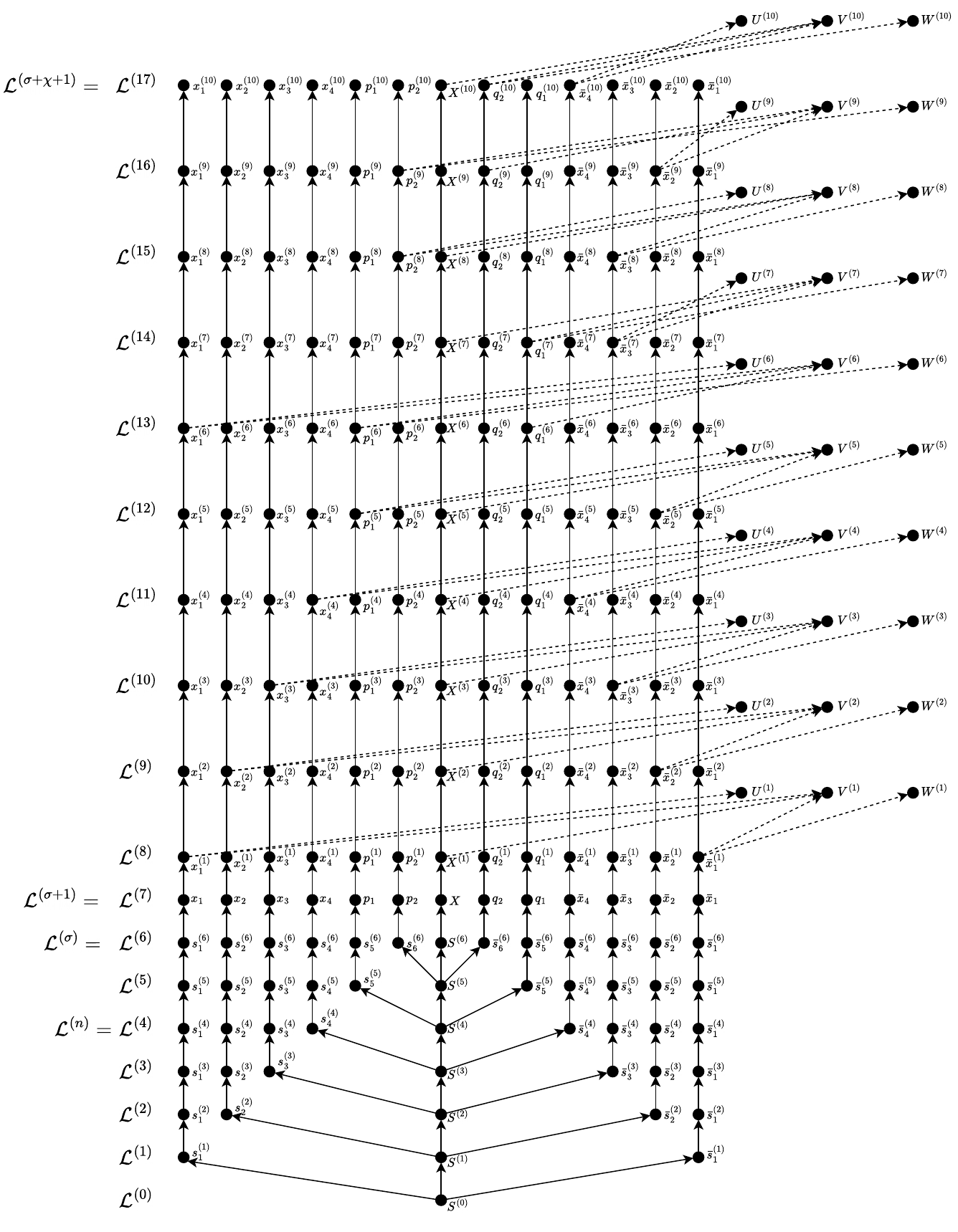}
\caption{The graph $G_{\Phi}$ for $\Phi = \NAE(x_1, x_2, \bar{x}_3) \land \NAE(\bar{x}_2, x_3, \bar{x}_4)$, where solid lines represent $0$-labeled arcs and dashed lines represent $1$-labeled arcs.}
\label{fig:graph}
\end{figure}

To prove \textsf{NP}-hardness, we give a polynomial-time reduction from NAE-3-SAT.
Given an instance $\Phi = \bigwedge_{j=1}^m \NAE(a_j, b_j, c_j)$ of NAE-3-SAT over variables $x_1, \dots, x_n$ with $n \geq 1$, we define $\sigma \coloneqq n+m$ and $\chi \coloneqq n+3m$ and construct a graph $G_\Phi = (\V, \E, \ell)$ as follows (see Fig.~\ref{fig:graph} for an example):
\begin{align}
\V &\coloneqq \{S^{(0)}\} \cup \bigcup_{d = 1}^{\sigma} \left(\{S^{(d)}\} \cup \bigcup_{i = 1}^d \{s_i^{(d)}, \bar{s}_i^{(d)}\}\right)\\
& \qquad\cup \{X\} \cup \{x_i, \bar{x}_i : i \in [n]\} \cup \{p_j, q_j : j \in [m]\}\\
& \qquad\cup \bigcup_{t = 1}^{\chi} \left( \{X^{(t)}\} \cup \bigcup_{i = 1}^n \{x_i^{(t)}, \bar{x}_i^{(t)}\} \cup \bigcup_{j = 1}^m \{p_j^{(t)}, q_j^{(t)}\} \cup \{U^{(t)}, V^{(t)}, W^{(t)}\} \right),\\
\E &\coloneqq \bigcup_{d = 1}^{\sigma} \left( \{S^{(d-1)} \xrightarrow{0} S^{(d)}, S^{(d-1)} \xrightarrow{0} s_d^{(d)}, S^{(d-1)} \xrightarrow{0} \bar{s}_d^{(d)}\} \cup \bigcup_{i = 1}^{d-1} \{s_i^{(d-1)} \xrightarrow{0} s_i^{(d)}, \bar{s}_i^{(d-1)} \xrightarrow{0} \bar{s}_i^{(d)}\} \right)\\
&\cup \{S^{(\sigma)} \xrightarrow{0} X\} \cup
\bigcup_{i = 1}^n \{s_i^{(\sigma)} \xrightarrow{0} x_i, \bar{s}_i^{(\sigma)} \xrightarrow{0} \bar{x}_i\} \cup
\bigcup_{j = 1}^m \{s_{n+j}^{(\sigma)} \xrightarrow{0} p_j, \bar{s}_{n+j}^{(\sigma)} \xrightarrow{0} q_j\}\\
&\cup \{X \xrightarrow{0} X^{(1)}\} \cup
\bigcup_{i = 1}^n \{ x_i \xrightarrow{0} x_i^{(1)}, \bar{x}_i \xrightarrow{0} \bar{x}_i^{(1)}\} \cup
\bigcup_{j = 1}^m \{p_j \xrightarrow{0} p_j^{(1)}, q_j \xrightarrow{0} q_j^{(1)}\}\\
&\cup \bigcup_{t = 1}^{\chi-1} \left(\{X^{(t)} \xrightarrow{0} X^{(t+1)}\} \cup \bigcup_{i = 1}^n \{x_i^{(t)} \xrightarrow{0} x_i^{(t+1)}, \bar{x}_i^{(t)} \xrightarrow{0} \bar{x}_i^{(t+1)}\} \cup 
\bigcup_{j = 1}^m \{p_j^{(t)} \xrightarrow{0} p_j^{(t+1)}, q_j^{(t)} \xrightarrow{0} q_j^{(t+1)}\} \right)\\ 
&\cup \bigcup_{i = 1}^n \mathcal{B}(x_i^{(i)}, X^{(i)}, \bar{x}_i^{(i)}; U^{(i)}, V^{(i)}, W^{(i)}) \label{eq:nfl98dtdcmoe}\\
&\cup \bigcup_{j = 1}^m \mathcal{B}(p_j^{(n+3j-2)}, X^{(n+3j-2)},\bar{b}_j^{(n+3j-2)}; U^{(n+3j-2)}, V^{(n+3j-2)}, W^{(n+3j-2)}) \label{eq:t1476qipq2nm}\\
& \cup \bigcup_{j = 1}^m \mathcal{B}(a_j^{(n+3j-1)}, q_j^{(n+3j-1)}, p_j^{(n+3j-1)}; U^{(n+3j-1)}, V^{(n+3j-1)}, W^{(n+3j-1)}) \label{eq:suqmqx98ly5j}\\
&\cup \bigcup_{j = 1}^m \mathcal{B}(c_j^{(n+3j)}, X^{(n+3j)}, q_j^{(n+3j)}; U^{(n+3j)}, V^{(n+3j)}, W^{(n+3j)}) \label{eq:zmq12tbvcv8i},
\end{align}
where
for $j \in [m]$ and $t \in [\chi]$,
the symbols $a_j^{(t)}, b_j^{(t)}, c_j^{(t)}, \bar{a}_j^{(t)}, \bar{b}_j^{(t)}, \bar{c}_j^{(t)}$ stand for the vertices with superscripts $(t)$ corresponding to the literals $a_j, b_j, c_j, \bar{a}_j, \bar{b}_j, \bar{c}_j$,
that is,
\begin{align}
a_j^{(t)}\coloneqq
\begin{cases}
x_i^{(t)} &\,\,\text{if}\,\, a_j = x_i,\\
\bar x_i^{(t)} &\,\,\text{if}\,\, a_j = \bar x_i,
\end{cases}
\quad
\bar{a}_j^{(t)}\coloneqq
\begin{cases}
\bar x_i^{(t)} &\,\,\text{if}\,\, a_j = x_i,\\
x_i^{(t)} &\,\,\text{if}\,\, a_j = \bar x_i,
\end{cases}
\end{align}
and $b_j^{(t)}, c_j^{(t)}, \bar{b}_j^{(t)}, \bar{c}_j^{(t)}$ are similarly defined.

For a non-negative integer $d$,
let $\mathcal{L}^{(d)}$ denote the set of vertices of $G_{\Phi}$ reachable from $S^{(0)}$ by traversing exactly $d$ $0$-labeled arcs.
Note that the literals $x_1, \dots, x_n, \bar{x}_1, \dots, \bar{x}_n$ are also vertices of $G_\Phi$, specifically belonging to the layer $\mathcal{L}^{(\sigma+1)}$.
For each $t \in [\chi]$, a copied vertex with superscript $(t)$ lies in the layer $\mathcal{L}^{(d)}$ at $d = \sigma + 1 + t$.
The gadget vertices $U^{(t)}, V^{(t)}, W^{(t)}$ lie outside these layers, since they are not reachable from $S^{(0)}$ by following $0$-labeled arcs.

The subgraph formed by the $0$-labeled arcs is a rooted tree on the vertices in $\mathcal{L}^{(0)} \cup \mathcal{L}^{(1)} \cup \cdots \cup \mathcal{L}^{(\sigma+\chi+1)}$.
In particular, every vertex in $\mathcal{L}^{(d)}$ for $d \geq 1$ has a unique parent in $\mathcal{L}^{(d-1)}$.
For $v \in \V$ and a non-negative integer $d$, let $\mathcal{D}^{(d)}(v)$ denote the set of all vertices in $\mathcal{L}^{(d)}$ reachable from $v$.

Figure~\ref{fig:graph} illustrates the graph $G_{\Phi}$ corresponding to an instance $\Phi = \NAE(x_1, x_2, \bar{x}_3) \land \NAE(\bar{x}_2, x_3, \bar{x}_4)$.

The constructed graph $G_\Phi$ satisfies the following properties:
\begin{itemize}
\item $G_\Phi$ is a DAG;
\item All vertices of $G_\Phi$ are reachable from the single source $S^{(0)}$;
\item The maximum out-degree is at most $3$;
\item The graph can be constructed in polynomial time since $|\V| = O((n+m)^2)$ and $|\E| = O((n+m)^2)$.
\end{itemize}

Therefore, to establish Theorem~\ref{thm:main}, it suffices to show that $G_\Phi$ is a Wheeler graph if and only if $\Phi$ is satisfiable. This equivalence is established by Lemmas~\ref{lem:soundness} and \ref{lem:completeness} below.

\begin{lemma}
\label{lem:soundness}
If $G_\Phi = (\V, \E, \ell)$ is a Wheeler graph, then $\Phi$ is satisfiable.
\end{lemma}

\begin{proof}
Let $\leq$ be a Wheeler order on $G_{\Phi}$.
Define a truth assignment $\tau \colon \{x_1, x_2, \ldots, x_n\} \to \{\mathsf{True}, \mathsf{False}\}$ as
\begin{align}
\tau(x_i) =
\begin{cases}
\mathsf{True} &\,\,\text{if}\,\, x_i < X,\\
\mathsf{False} &\,\,\text{if}\,\, x_i > X
\end{cases}
\label{eq:jx3dbtsusqwz}
\end{align}
for $i \in [n]$.
We claim that $\tau$ satisfies $\Phi$.

Lemma~\ref{lem:gadget} implies that the order $\le$ satisfies the following betweenness constraints since $G_{\Phi}$ contains the gadgets~\eqref{eq:nfl98dtdcmoe}, \eqref{eq:t1476qipq2nm}, \eqref{eq:suqmqx98ly5j}, and \eqref{eq:zmq12tbvcv8i}:
\begin{align}
B(x_i^{(i)}, X^{(i)}, \bar{x}_i^{(i)}) &\,\,\text{for all}\,\, i \in [n], \label{eq:const-1} \\
B(p_j^{(n+3j-2)}, X^{(n+3j-2)}, \bar{b}_j^{(n+3j-2)}) &\,\,\text{for all}\,\, j \in [m], \label{eq:const-2}\\
B(a_j^{(n+3j-1)}, q_j^{(n+3j-1)}, p_j^{(n+3j-1)}) &\,\,\text{for all}\,\, j \in [m], \label{eq:const-3}\\
B(c_j^{(n+3j)}, X^{(n+3j)}, q_j^{(n+3j)}) &\,\,\text{for all}\,\, j \in [m]. \label{eq:const-4}
\end{align}

For any $d \in \{\sigma+2, \sigma+3, \ldots, \sigma+\chi+1\}$ and distinct $u, v \in \mathcal{L}^{(d)}$,
there exist arcs $\hat{u} \xrightarrow{0} u$ and $\hat{v} \xrightarrow{0} v$ from distinct parent vertices $\hat{u}, \hat{v} \in \mathcal{L}^{(d-1)}$,
so that Definition~\ref{def:wheeler}~(b2) ensures that
the relative order is preserved across layers:
\begin{align}
\hat{u} < \hat{v} \iff u < v.
\end{align}
Thus, the relative order established in $\mathcal{L}^{(\sigma+1)}$ is preserved across
all subsequent layers $\mathcal{L}^{(\sigma+2)}, \allowbreak \mathcal{L}^{(\sigma+3)}, \ldots, \mathcal{L}^{(\sigma+\chi+1)}$.
Consequently, the constraints~\eqref{eq:const-1}, \eqref{eq:const-2}, \eqref{eq:const-3}, and \eqref{eq:const-4} propagate back to the layer $\mathcal{L}^{(\sigma+1)}$, implying that the following constraints hold:
\begin{align}
B(x_i, X, \bar{x}_i) &\,\,\text{for all}\,\, i \in [n], \label{eq:klhrvcom4bp7} \\
B(p_j, X, \bar{b}_j) &\,\,\text{for all}\,\, j \in [m], \label{eq:ot82bk1uotxs}\\
B(a_j, q_j, p_j) &\,\,\text{for all}\,\, j \in [m], \label{eq:jxx6f60hjpwm}\\
B(c_j, X, q_j) &\,\,\text{for all}\,\, j \in [m]. \label{eq:4e0xzd4nctk}
\end{align}
The condition \eqref{eq:klhrvcom4bp7} forces $\bar{x}_i$ to lie on the opposite side of $X$ relative to $x_i$ (i.e., either $x_i < X < \bar{x}_i$ or $\bar{x}_i < X < x_i$).
Thus, \eqref{eq:jx3dbtsusqwz} yields
\begin{align}
\tau(x_i) = \mathsf{True} &\iff x_i < X < \bar{x}_i, \label{eq:55tgbod29cqi-1}\\
\tau(x_i) = \mathsf{False} &\iff \bar{x}_i < X < x_i \label{eq:55tgbod29cqi-2}
\end{align}
for any $i \in [n]$.

For $j \in [m]$, we show by contradiction that the clause $\NAE(a_j, b_j, c_j)$ of $\Phi$ is satisfied as follows.
Assume that all three literals of this clause are $\mathsf{True}$.
Then \eqref{eq:55tgbod29cqi-1} and \eqref{eq:55tgbod29cqi-2} directly imply
\begin{align}
a_j &< X < \bar{a}_j, \label{eq:gjzomq68b056-1}\\
b_j &< X < \bar{b}_j, \label{eq:gjzomq68b056-2}\\
c_j &< X < \bar{c}_j. \label{eq:gjzomq68b056-3}
\end{align}
The right inequality in \eqref{eq:gjzomq68b056-2} and the constraint \eqref{eq:ot82bk1uotxs} imply
\begin{align}
p_j < X. \label{eq:ds5htkk777kd}
\end{align}
Also, the left inequality in \eqref{eq:gjzomq68b056-3} and the constraint \eqref{eq:4e0xzd4nctk} imply
\begin{align}
X < q_j. \label{eq:xqp9perr78g1}
\end{align}
Combining the constraint \eqref{eq:jxx6f60hjpwm} with \eqref{eq:ds5htkk777kd} and \eqref{eq:xqp9perr78g1},
we obtain $p_j < X < q_j < a_j$, which contradicts the left inequality in \eqref{eq:gjzomq68b056-1}.
The case where all three literals of the clause are $\mathsf{False}$ also leads to a contradiction symmetrically.

Consequently, $\tau$ satisfies all clauses, so that $\Phi$ is satisfiable.
\end{proof}

\begin{lemma}
\label{lem:completeness}
If $\Phi$ is satisfiable, then $G_\Phi$ is a Wheeler graph.
\end{lemma}

\begin{proof}
Given a satisfying truth assignment $\tau \colon \{x_1, x_2, \ldots, x_n\} \to \{\mathsf{True}, \mathsf{False}\}$,
we construct an order $\leq$ on $\V$ according to the following rules (i)--(iii).
\begin{enumerate}[(i)]
\item Order the vertex sets globally as follows:
\begin{align}
&\mathcal{L}^{(0)} < \mathcal{L}^{(1)} < \cdots < \mathcal{L}^{(\sigma+\chi+1)} <\\
& \quad \{U^{(1)}, V^{(1)}, W^{(1)}\} < \{U^{(2)}, V^{(2)}, W^{(2)}\} < \cdots < \{U^{(\chi)}, V^{(\chi)}, W^{(\chi)}\},
\end{align}
where for vertex sets $\mathcal{A}$ and $\mathcal{B}$, the notation $\mathcal{A} < \mathcal{B}$ means that $u < v$ holds for every $u \in \mathcal{A}$ and $v \in \mathcal{B}$.

\item For every $d \in [\sigma+\chi+1]$, the relative order in $\mathcal{L}^{(d)}$ is defined inductively in increasing order of $d$ by the following rules (A) and (B).
\begin{enumerate}[(A)]
\item If two distinct vertices $u, v \in \mathcal{L}^{(d)}$ have different parents $\hat{u}, \hat{v} \in \mathcal{L}^{(d-1)}$, then their relative order is inherited from their parents, i.e., $u < v \iff \hat{u} < \hat{v}$.
\item The relative order in $\{s_d^{(d)}, \bar{s}_d^{(d)}, S^{(d)}\}$ for $d \in [\sigma]$, whose vertices share the same parent $S^{(d-1)}$, is determined according to the following rules (I) and (II).

\begin{enumerate}[(I)]
\item The case $1 \leq d \le n$:
if $\tau(x_d) = \mathsf{True}$, then set the order to $s_d^{(d)} < S^{(d)} < \bar{s}_d^{(d)}$;
if $\tau(x_d) = \mathsf{False}$, then set the order to $\bar{s}_d^{(d)} < S^{(d)} < s_d^{(d)}$.
\item The case $n+1 \le d \le \sigma$:
select the ordering according to 
the column ``Order on $\{s_d^{(d)}, \bar{s}_d^{(d)}, S^{(d)}\}$'' in Table~\ref{tab:clause_orders} depending on the truth values of $a_j, b_j, c_j$,
where $j \coloneqq d-n$.
\end{enumerate}
\end{enumerate}
Note that applying the rule (A) $d'-d$ times shows that
for any $d, d' \in [\sigma+\chi+1]$ with $d < d'$ and $u, v \in \mathcal{L}^{(d)}$,
\begin{align}
u < v \implies \mathcal{D}^{(d')}(u) < \mathcal{D}^{(d')}(v).
\label{eq:0gdzgnihfqt6}
\end{align}

\item
For every $t \in [\chi]$,
we determine the relative order on $\{U^{(t)}, V^{(t)}, W^{(t)}\}$
so that Definition~\ref{def:wheeler}~(b1) and~(b2) hold for all pairs among the five arcs of the betweenness gadget containing $U^{(t)}, V^{(t)}, W^{(t)}$.
\end{enumerate}

\begin{table}[htbp]
\centering
\caption{Order of the corresponding vertices based on truth values of literals, where $j \coloneqq d-n$,
and $\mathsf{T}$ (resp.~$\mathsf{F}$) represents $\mathsf{True}$ (resp.~$\mathsf{False}$).}
\label{tab:clause_orders}
\begin{tabular}{ccccc}
\toprule
$a_j$ & $b_j$ & $c_j$ & Order on $\{s_d^{(d)}, \bar{s}_d^{(d)}, S^{(d)}\}$  & Order on the corresponding vertices in $\mathcal{L}^{(\sigma+1)}$\\
\midrule
$\mathsf{T}$ & $\mathsf{T}$ & $\mathsf{F}$ & $\bar{s}_d^{(d)} < s_d^{(d)} < S^{(d)}$ & $\{a_j, b_j, \bar{c}_j\} < \{q_j\} < \{p_j\} < \{X\} < \{\bar{a}_j, \bar{b}_j, c_j\}$ \\
$\mathsf{T}$ & $\mathsf{F}$ & $\mathsf{T}$ & $S^{(d)} < \bar{s}_d^{(d)} < s_d^{(d)}$ & $\{a_j, \bar{b}_j, c_j\} < \{X\} < \{q_j\} < \{p_j\} < \{\bar{a}_j, b_j, \bar{c}_j\}$ \\
$\mathsf{F}$ & $\mathsf{T}$ & $\mathsf{T}$ & $s_d^{(d)} < S^{(d)} < \bar{s}_d^{(d)}$ & $\{\bar{a}_j, b_j, c_j\} < \{p_j\} < \{X\} < \{q_j\} < \{a_j, \bar{b}_j, \bar{c}_j\}$ \\
$\mathsf{T}$ & $\mathsf{F}$ & $\mathsf{F}$ & $\bar{s}_d^{(d)} < S^{(d)} < s_d^{(d)}$ & $\{a_j, \bar{b}_j, \bar{c}_j\} < \{q_j\} < \{X\} < \{p_j\} < \{\bar{a}_j, b_j, c_j\}$ \\
$\mathsf{F}$ & $\mathsf{T}$ & $\mathsf{F}$ & $s_d^{(d)} < \bar{s}_d^{(d)} < S^{(d)}$ & $\{\bar{a}_j, b_j, \bar{c}_j\} < \{p_j\} < \{q_j\} < \{X\} < \{a_j, \bar{b}_j, c_j\}$ \\
$\mathsf{F}$ & $\mathsf{F}$ & $\mathsf{T}$ & $S^{(d)} < s_d^{(d)} < \bar{s}_d^{(d)}$ & $\{\bar{a}_j, \bar{b}_j, c_j\} < \{X\} < \{p_j\} < \{q_j\} < \{a_j, b_j, \bar{c}_j\}$ \\
\bottomrule
\end{tabular}
\end{table}

The existence of the relative order required by the rule (iii) is guaranteed by Lemma~\ref{lem:gadget}
because the rules (i) and (ii) guarantee that the betweenness constraints
\begin{align}
B(x_i^{(i)}, X^{(i)}, \bar{x}_i^{(i)}) &\,\,\text{for all}\,\, i \in [n], \label{eq:rl8jkqdp74kb1}\\
B(p_j^{(n+3j-2)}, X^{(n+3j-2)}, \bar{b}_j^{(n+3j-2)}) &\,\,\text{for all}\,\, j \in [m], \label{eq:rl8jkqdp74kb2}\\
B(a_j^{(n+3j-1)}, q_j^{(n+3j-1)}, p_j^{(n+3j-1)}) &\,\,\text{for all}\,\, j \in [m], \label{eq:rl8jkqdp74kb3}\\
B(c_j^{(n+3j)}, X^{(n+3j)}, q_j^{(n+3j)}) &\,\,\text{for all}\,\, j \in [m] \label{eq:rl8jkqdp74kb4}
\end{align}
hold as shown below.

For $i \in [n]$, if $\tau(x_i) = \mathsf{True}$, then the rule (I) places $s_i^{(i)} < S^{(i)} < \bar{s}_i^{(i)}$, which implies
\begin{align}
\{x_i\} = \mathcal{D}^{(\sigma+1)}(s_i^{(i)}) < \mathcal{D}^{(\sigma+1)}(S^{(i)}) < \mathcal{D}^{(\sigma+1)}(\bar{s}_i^{(i)}) = \{\bar{x}_i\}
\end{align}
by \eqref{eq:0gdzgnihfqt6}.
Similarly, if $\tau(x_i) = \mathsf{False}$, then
\begin{align}
\{\bar{x}_i\} = \mathcal{D}^{(\sigma+1)}(\bar{s}_i^{(i)}) < \mathcal{D}^{(\sigma+1)}(S^{(i)}) < \mathcal{D}^{(\sigma+1)}(s_i^{(i)}) = \{x_i\}.
\end{align}
Since $\{X, p_1, p_2, \ldots, p_m, q_1, q_2, \ldots, q_m\} \subseteq \mathcal{D}^{(\sigma+1)}(S^{(i)})$, 
the relative order in $\mathcal{L}^{(\sigma+1)}$ satisfies
\begin{align}
\mathcal{T} < \{X, p_1, p_2, \ldots, p_m, q_1, q_2, \ldots, q_m\} < \mathcal{F}, \label{eq:xel0n91xa8w5}
\end{align}
where $\mathcal{T}$ (resp.~$\mathcal{F}$) denotes the set of vertices in $\mathcal{L}^{(\sigma+1)}$ corresponding to $\mathsf{True}$ (resp.~$\mathsf{False}$) literals  under $\tau$.
This directly implies that the betweenness constraints hold:
\begin{align}
B(x_i, X, \bar{x}_i) \,\,\text{for all}\,\, i \in [n] \label{eq:oxrbfyirp0la}.
\end{align}

By \eqref{eq:xel0n91xa8w5} and the rule (II), for each clause $\NAE(a_j, b_j, c_j)$, the relative order of the corresponding vertices in layer $\mathcal{L}^{(\sigma+1)}$ matches the entry in the rightmost column of Table~\ref{tab:clause_orders}, according to the truth values of $a_j, b_j$, and $c_j$.
For example, in the case of the first row in the table,
the relative order $\bar{s}_d^{(d)} < s_d^{(d)} < S^{(d)}$ set by the rule (II) implies
\begin{align}
\mathcal{D}^{(\sigma+1)}(\bar{s}_d^{(d)}) < \mathcal{D}^{(\sigma+1)}(s_d^{(d)}) < \mathcal{D}^{(\sigma+1)}(S^{(d)})
\end{align}
by \eqref{eq:0gdzgnihfqt6}.
Since $\mathcal{D}^{(\sigma+1)}(\bar{s}_d^{(d)}) = \{q_j\}$, 
$\mathcal{D}^{(\sigma+1)}(s_d^{(d)}) = \{p_j\}$, and $\mathcal{D}^{(\sigma+1)}(S^{(d)}) \supseteq \{X\}$, we obtain $q_j < p_j < X$.
Combining this inequality and \eqref{eq:xel0n91xa8w5} yields the entry in the rightmost column.
The other rows in the table are similarly confirmed.

Direct case inspection confirms that every row in Table~\ref{tab:clause_orders} satisfies the betweenness constraints
\begin{align}
B(p_j, X, \bar{b}_j), B(a_j, q_j, p_j), B(c_j, X, q_j) &\,\,\text{for all}\,\, j \in [m].
\label{eq:xwp376914doo}
\end{align}
By \eqref{eq:0gdzgnihfqt6}, the relative order in $\mathcal{L}^{(\sigma+1)}$ is preserved in all subsequent layers $\mathcal{L}^{(\sigma+2)}, \mathcal{L}^{(\sigma+3)}, \ldots, \allowbreak \mathcal{L}^{(\sigma+\chi+1)}$, thereby ensuring that \eqref{eq:oxrbfyirp0la} and \eqref{eq:xwp376914doo} imply the desired conditions \eqref{eq:rl8jkqdp74kb1}, \eqref{eq:rl8jkqdp74kb2}, \eqref{eq:rl8jkqdp74kb3}, and \eqref{eq:rl8jkqdp74kb4}.

The order $\leq$ defined above is indeed a Wheeler order; that is, it satisfies the conditions (a), (b1), and (b2) of Definition~\ref{def:wheeler} as follows.
\begin{itemize}
\item[(a)] The only vertex of in-degree zero is $S^{(0)}$, and the rule~(i) places it first.
Hence, Definition~\ref{def:wheeler} (a) holds.

\item[(b1)]
Every $0$-labeled arc ends at a vertex of $\mathcal{L}^{(d)}$ for some $d \in [\sigma+\chi+1]$, whereas every $1$-labeled arc ends at a vertex of $\{U^{(t)}, V^{(t)}, W^{(t)}\}$ for some $t \in [\chi]$.
The above rule (i) places all the former vertices before all the latter vertices,
so that Definition~\ref{def:wheeler} (b1) holds for every pair of arcs.

\item[(b2)]
Choose $e = (u, v) \in \E$ and $e' = (u', v') \in \E$ with $\ell(e) = \ell(e')$ and $u < u'$.
Suppose $u \in \mathcal{L}^{(d)}$ and $u' \in \mathcal{L}^{(d')}$.
We show that $v \leq v'$ as follows.

\begin{itemize}
\item The case $\ell(e) = \ell(e') = 0$:
If $d = d'$ (i.e., $u$ and $u'$ belong to the same layer), then the rule (A) orders their children in the same order as $u$ and $u'$, giving $v < v'$.
If $d \neq d'$ (i.e., $u$ and $u'$ belong to different layers), then the rule (i) implies $d < d'$ since $u < u'$.
Their children $v$ and $v'$ belong to $\mathcal{L}^{(d+1)}$ and $\mathcal{L}^{(d'+1)}$, respectively, so again $v < v'$ by the rule (i).

\item The case $\ell(e) = \ell(e') = 1$:
If $d = d'$ (i.e., $u$ and $u'$ belong to the same betweenness gadget), then $v \leq v'$ follows directly from the rule (iii).
If $d \neq d'$ (i.e., $u$ and $u'$ belong to distinct betweenness gadgets), then the rule (i) implies $d < d'$ since $u < u'$.
The vertices $v$ and $v'$ belong to $\{U^{(k)}, V^{(k)}, W^{(k)}\}$ and $\{U^{(k')}, V^{(k')}, W^{(k')}\}$, respectively, where
$k \coloneqq d - (\sigma+1)$ and $k' \coloneqq d' - (\sigma+1)$.
Thus, the rule (i) implies  $v \in \{U^{(k)}, V^{(k)}, W^{(k)}\} < \{U^{(k')}, V^{(k')}, W^{(k')}\} \owns v'$, giving $v < v'$.
\end{itemize}
\end{itemize}

Consequently, $\le$ constitutes a valid Wheeler order on $G_{\Phi}$.
\end{proof}

\section*{Acknowledgements}
The proof of Theorem \ref{thm:main} was obtained during a session with GPT-6 Astra.

\bibliographystyle{plain}
\bibliography{arxiv}

\end{document}